\documentstyle [12pt,epsfig]{article} 
\makeatletter
\@addtoreset{equation}{section}
\makeatother

\newcommand{\ba}{\begin{eqnarray}}
\newcommand{\ea}{\end{eqnarray}}

\begin{document}
\newcommand{\BS}{\bigskip}
\newcommand{\SECTION}[1]{\BS{\large\section{\bf #1}}}
\newcommand{\SUBSECTION}[1]{\BS{\large\subsection{\bf #1}}}
\newcommand{\SUBSUBSECTION}[1]{\BS{\large\subsubsection{\bf #1}}}

\begin{titlepage}
\begin{center}
\vspace*{2cm}
{\large \bf Derivation of the Lorentz Force Law and the Magnetic Field Concept using
  an Invariant Formulation of the Lorentz Transformation}
\vspace*{1.5cm}
\end{center}
\begin{center}
{\bf J.H.Field }
\end{center}
\begin{center}
{ 
D\'{e}partement de Physique Nucl\'{e}aire et Corpusculaire
 Universit\'{e} de Gen\`{e}ve . 24, quai Ernest-Ansermet
 CH-1211 Gen\`{e}ve 4.}
\end{center}
\begin{center}
{e-mail; john.field@cern.ch}
\end{center}
\vspace*{2cm}
\begin{abstract}
 It is demonstrated how the right hand sides of the Lorentz Transformation
  equations may be written, in a Lorentz invariant manner, as 4--vector
  scalar products. The formalism is shown to provide a short derivation,
   in which the 4--vector electromagnetic potential plays a crucial role, of the 
  Lorentz force law of classical electrodynamics, and the conventional
   definition of the magnetic field in terms spatial derivatives of
  the 4--vector potential. The time component
  of the relativistic generalisation of the Lorentz force law is discussed. An important
  physical distinction between the space-time and energy-momentum 4--vectors is
  also pointed out.
\end{abstract}
\vspace*{1cm}
{\it Keywords};  Special Relativity, Classical Electrodynamics.
\newline
\vspace*{1cm}
\end{titlepage}

\SECTION{\bf{Introduction}}

  Numerous examples exist in the literature of the derivation of electrodynamical
 equations from simpler physical hypotheses. In Einstein's original paper on
  Special Relativity~\cite{Einstein}, the Lorentz force law was derived by
  performing a Lorentz transformation of the electromagnetic fields and the space-time
 coordinates from the rest frame of an electron (where only electrostatic forces act)
 to the laboratory system where the electron is in motion and so also subjected
  to magnetic forces. A similar demonstration was given by Schwartz~\cite{Schwartz}
 who also showed how the electrodynamical Maxwell equations can be derived from the
  Gauss laws of electrostatics and magnetostatics by exploiting the 4-vector character
 of the electromagnetic current and the symmetry properties of the electromagnetic
  field tensor. The same type of derivation of electrodynamic Maxwell equations from
  the electrostatic and magnetostatic ones has recently been performed by the 
  present author on the basis of `space-time exchange symmetry'~\cite{JHF}.
  Frisch and Wilets~\cite{FW} discussed the derivation of Maxwell's equations
  and the Lorentz force law by application of relativistic transforms to the
   electrostatic Gauss law. Dyson~\cite{Dyson} published a proof, due originally to 
  Feynman, of the Faraday-Lenz law of induction, based on Newton's Second Law
  and the quantum commutation relations of position and momentum,
  that excited considerable interest and a flurry of comments and publications~\cite{Dombey,
   Brehme,Anderson,Farquhar,Tanimura,VF}
  about a decade ago. Landau and Lifshitz~\cite{LLCTF} presented a derivation
  of Amp\`{e}re's Law from the electrodynamic Lagrangian, using the Principle of
  Least Action. By relativistic transformation of the Coulomb force from the rest frame
  of a charge to another inertial system in relative motion, Lorrain, Corson and Lorrain~\cite{LCL}
 derived both the Biot-Savart law, for the magnetic field generated by a moving charge,
  and the Lorentz force law.
  \par In many text books on classical electrodynamics the question of what are the
   fundamental physical hypotheses underlying the subject, as distinct from purely
 mathematical developments of these hypotheses, used to derive predictions, is not discussed
 in any detail. Indeed, it may even be stated that it is futile to address the question
  at all. For example, Jackson~\cite{Jack1} states:
  \par {\sl At present it is popular in undergraduate texts and elsewhere to attempt to derive
   magnetic fields and even Maxwell equations from Coulomb's law of electrostatics and
 the theory of Special Relativity. It should immediately obvious that, without additional
 assumptions, this is impossible.'}
  \par This is, perhaps, a true statement. However, if the additional assumptions are weak ones,
   the derivation may still be a worthwhile exercise.    
    In fact, in the case of Maxwell's equations, as shown
   in References~\cite{Schwartz,JHF}, the `additional assumptions' are merely the formal definitions
   of the electric and magnetic fields in terms of the space--time derivatives of
   the 4--vector potential~\cite{JHF1}. In the case of the
    derivation of the Lorentz force equation given below, not even the latter assumption is required,
    as the magnetic field definition appears naturally in the course of the derivation.
  \par In the chapter on `The Electromagnetic Field' in Misner Thorne and Wheeler's book
    `Gravitation'~\cite{MTW} can be found the following statement:
   \par  {\sl Here and elsewhere in science, as stressed not least by Henri Poincar\'{e}, that view 
   is out of date which used to say, ``Define your terms before you proceed''. All the laws
 and theories of physics, including the Lorentz force law, have this deep and subtle 
  chracter, that they both define the concepts they use (here $\vec{B}$ and $\vec{E}$) and make 
   statements about these concepts. Contrariwise, the absence of some body of theory, law
  and principle deprives one of the means properly to define or even use concepts. Any forward step
    in human knowlege is truly creative in this sense: that theory concept, law, and measurement
     ---forever inseperable---are born into the world in union.}

    \par I do not agree that the electric and magnetic fields are the fundamental concepts 
     of electromagnetism, or that the Lorentz force law cannot be derived from
    simpler and more fundamental concepts, but must be `swallowed whole', as this passage 
    suggests. As demonstrated in References~\cite{Schwartz,JHF} where the
    electrodynamic and magnetodynamic Maxwell equations are derived from those of
    electrostatics and magnetostatics, a more economical description of classical electromagentism
    is provided by the 4--vector potential. Another example
  of this is provided by the derivation of the Lorentz force law presented in the present paper.
   The discussion of electrodynamics in Reference~\cite{MTW} is couched entirely in terms of
  the electromagnetic field tensor, $F^{\mu \nu}$, and the electric and magnetic fields which,
   like the Lorentz force law and Maxwell's equations, are `parachuted' into 
   the exposition without any proof or any discussion of their interrelatedness. The 4--vector
    potential is 
   introduced only in the next-but-last exercise at the end of the
   chapter. After the derivation of the Lorentz force law in Section 3 below, a comparison
   will be made with the
    treatment of the law in References~\cite{Schwartz,Jack1,MTW}.
   \par The present paper introduces, in the following Section, the idea of an `invariant
   formulation' of the Lorentz Transformation (LT)~\cite{Fahnline}. It will be shown that
    the RHS of the LT equations of space and time can be written as 4-vector scalar products, so
   that the transformed 4-vector components are themselves Lorentz invariant
    quantities. Consideration of particular length and time interval measurements demonstrates 
    that this is a physically meaningful concept. It is pointed out that, whereas space
   and time intervals are, in general, physically independent physical quantities,
   this is not the case for the space and time components of the energy-momentum
   4-vector. In Section 3, a derivation of the Lorentz force law,
   and the associated magnetic field concept, is given, based on the
  invariant formulation of the LT. The derivation is very short, the only initial
   hypothesis being the usual definition of the electric field in terms of the
   4-vector potential, which, in fact, is also uniquely specified by requiring the
   definition to be a covariant one.
   In Section 4 the time component of Newton's Second Law in
   electrodynamics, obtained by applying space-time exchange symmetry~\cite{JHF}
   to the Lorentz force law, is discussed.
     \par Throughout this paper it is assumed that the electromagnetic field  
     constitutes, together with the moving charge, a conservative system; i.e.
    effects of radiation, due to the acceleration of the charge, are neglected

 \SECTION{\bf{Invariant Formulation of the Lorentz Transformation}}
  The space-time LT equations between two inertial frames S and S',
  written in a space-time symmetric manner, are:
  \begin{eqnarray}
  x' &=& \gamma (x-\beta x^0) \\
  y' &=& y  \\
  z' &=& z  \\ 
  x'^0 &=& \gamma(x^0-\beta x)
  \end{eqnarray}  
 The frame S' moves with velocity, $v$, relative to S, along the common x-axis of S and S'.
  $\beta$ and $\gamma$ are the usual relativistic parameters:
 \begin{eqnarray}
 \beta &\equiv& \frac{v}{c}    \\
 \gamma &\equiv& \frac{1}{\sqrt{1-\beta^2}}
 \end{eqnarray}  
  where $c$ is the speed of light, and 
  \begin{equation}
   x^0 \equiv c t
  \end{equation}
   where $t$ is the time recorded by an observer at rest in S. Clocks 
  in S and S' are synchronised, i.e., $t = t' = 0$, when the origins of the spatial
  cordinates of S and S' coincide.
   \par Eqns(2.1)-(2.4) give the relation between space and time intervals
     $\Delta x$, $\Delta x^0 = c \Delta t$ as observed in the two frames:
   \begin{eqnarray}
 \Delta x' &=& \gamma ( \Delta x-\beta  \Delta x^0) \\
  \Delta y' &=& \Delta y  \\
  \Delta z' &=& \Delta z  \\ 
  \Delta x'^0 &=& \gamma(\Delta x^0-\beta \Delta x)
  \end{eqnarray}  
   Each interval can be interpreted as the result of a particular measurement
  performed in the appropriate frame. For example, $\Delta x$ may correspond to
  the measurement of the distance between two points lying along the x-axis at a fixed 
  time in S. In virtue of this, it may be identified with the space-like invariant
   interval, $S_x$, where:
  \begin{equation}
    S_x \equiv \sqrt{ -(\Delta x^0)^2+\Delta x^2} = \Delta x 
  \end{equation}
   since, for the measurement procedure just described, $\Delta x^0 = 0$. Notice that
   $\Delta x$ is not {\it necessarily} defined in terms of such a measurement. If, following
   Einstein~\cite{Einstein}, the interval $\Delta x$ is associated with the length, $\ell$, of a measuring
   rod at rest in S and lying parallel to the x-axis, measurements of the ends of
   the rod can be made at arbitarily different times in S. The same result 
   $\ell = \Delta x$ will be found for the length of the rod, but the corresponding 
   invariant interval, $S_x$, as defined by Eqn(2.12) will be different in each case.
   Similarly, $\Delta x^0$ may be identified with the time-like invariant interval
   corresponding to successive observations of a clock at a fixed position
   (i.e. $\Delta x = 0$) in S:
   \begin{equation}
    S_0 \equiv  \sqrt{(\Delta x^0)^2-\Delta x^2} = \Delta x^0 
  \end{equation}
    The interval $\Delta x^0$ could also be measured by observing the difference of the
   times recorded by a local clock and another, synchronised, one located at a different
   position in S, after a suitable correction for light propagation time delay.
   Each such pair of clocks would yield the same value, $\Delta x^0$, for the time 
   difference between two events in S, but with different values of the invariant
   interval defined by Eqn(2.13).
   \par In virtue of Eqns(2.12) and (2.13) the LT equations (2.8) and (2.11) 
    may be written the following {\it invariant} form:
   \begin{eqnarray}
  S'_x & = & -\bar{U}(\beta)\cdot S \\
  S'_0 & = & U(\beta)\cdot S
  \end{eqnarray} 
   where the following 4--vectors have been introduced:
   \begin{eqnarray}
  S & \equiv & (S_0; S_x, 0, 0) = ( \Delta x^0; \Delta x, 0, 0) \\
  U(\beta) & \equiv & ( \gamma ; \gamma \beta, 0, 0) \\
 \bar{U}(\beta) &  \equiv & ( \gamma\beta ;\gamma, 0, 0) 
  \end{eqnarray} 
   The time-like 4-vector, $U$, is equal to $V/c$, where $V$  is the usual 4--vector velocity,
 whereas the space-like 4--vector, $\bar{U}$, is `orthogonal to $U$ in
  four dimensions':
  \begin{equation}
   U(\beta) \cdot \bar{U}(\beta) = 0
  \end{equation}
  Since the RHS of (2.14) and (2.15) are 4--vector scalar products,
  $S'_x$ and $S'_0$ are manifestly Lorentz invariant quantites. These 
  4--vector components may be defined, in terms of specific space-time
   measurements,  by equations similar to (2.12) and (2.13) 
  in the frame S'. Note that the 4--vectors $S$ and $S'$ are `doubly covariant' 
  in the sense that $S \cdot S$ and  $S' \cdot S'$ are `doubly invariant'
  quantities whose spatial and temporal terms are, individually, 
  Lorentz invariant:   
   \begin{equation}
  S \cdot S = S_0^2-S_x^2 = S'\cdot S' =  (S'_0)^2-(S'_x)^2     
   \end{equation}
  Every term in Eqn(2.20) remains invariant if the spatial and temporal
  intervals described above are observed from a third inertial frame
   S'' moving along the x-axis relative to both S and S'. This follows from 
   the manifest Lorentz invariance of the RHS of Eqn(2.14) and (2.15)
   and their inverses:
   \begin{eqnarray}
  S_x & = & -\bar{U}(-\beta)\cdot S' \\
  S_0 & = & U(-\beta)\cdot S'
  \end{eqnarray}
  \par  Since the LT Eqns(2.1) and (2.4) are valid for any 4--vector, $W$, it follows
   that: 
   \begin{eqnarray}
  W'_x & = & -\bar{U}(\beta)\cdot W \\
  W'_0 & = & U(\beta)\cdot W
  \end{eqnarray} 
 Again, $W'_x$ and $W'_0$ are manifestly Lorentz invariant.
  An interesting special case is the energy-momentum 4--vector, $P$, of a physical 
  object of mass, $m$. Here the `doubly invariant' quantity analagous to $S \cdot S$
 in Eqn(2.20) is equal to $m^2c^2$. Choosing the  x-axis parallel
   to $\vec{p}$ and $\beta$ to correspond to the object's velocity, so that
   S' is the object's proper frame, and since $ P \equiv mc U(\beta)$, 
   Eqns(2.23) and (2.24) yield, for this special case: 
   \begin{eqnarray}
  P'_x & = & - mc \bar{U}(\beta)\cdot U(\beta) = 0 \\
  P'_0 & = &  m c U(\beta)\cdot U(\beta)  = mc
  \end{eqnarray} 
    Since the Lorentz transformation is determined by the single parameter, $\beta$, then
   it follows from Eqns(2.25) and (2.26) that,
   unlike in the case of the space and time intervals in Eqns(2.8) and (2.11),
   the spatial and temporal components of the energy momentum
   4--vector, in an arbitary inertial frame, are not independent. In fact, $P_0$ is  determined 
   in terms of $P_x$ and $m$ by the relation, that follows from the
    inverse of Eqns(2.25) and (2.26): $P_0 = \sqrt{P_x^2 + m^2c^2}$.
   Thus, although the LT equations for the space-time and
   energy-momentum 4--vectors are mathematically identical, the physical
   interpretation of the transformed quantities is quite different
   in the two cases. 
   \par  The LT equation, (2.24), for the electromagnetic 4--vector potential, $A$, is found to
     play a crucial
    role in the derivation of the Lorentz force law presented in the following
    Section.

\SECTION{\bf{Derivation of the Lorentz force law and the Magnetic Field}}  
    In electrostatics, the electric field, $\vec{E}$, is customarily written in terms
   of the electrostatic potential, $\phi$, according to the equation $\vec{E} = -\vec{\nabla} \phi$.
   The potential at a distance, $r$, from a point charge, $Q$, is given by Coulomb's
   law $\phi(r) = Q/r$. This, together with the equation
   $\vec{F} =q \vec{E}$, defining the force, $\vec{F}$, exerted on a charge, $q$, by the
   electric field, completes the specification of the dynamical basis of
   classical electromagnetism.
   \par It remains to generalise the above equation relating the 
    electric field to the electrostatic potential in a manner consistent with special
    relativity. In relativistic notation~\cite{Defns}, the electric field is related to the
   potential by the equation: $E^i = \partial^i A^0$, where $\phi$ is identified with 
   the time component, $A^0$, of the 4--vector electromagnetic potential ($A^0$;$\vec{A}$).
    In order to respect special relativity the electric field must be defined in a covariant
     manner, i.e. in the same way in all inertial frames. The electrostatic law may
   be generalised in two ways:
 \begin{equation}
   E^i \rightarrow E^i_{\pm} \equiv \partial^i A^0 \pm\partial^0 A^i
   \end{equation}
  This equation shows the only possiblities to define the electric field in a way that respects the
  symmetry with respect to the exchange of space and time coordinates that is a general
  property of all special relativistic laws~\cite{JHF}. Choosing $i = 1$ in Eqn(3.1) and transforming
   all quantities on the RHS into the S' frame, by use of the inverses of Eqns(2.1) and (2.4),
   leads to the following expressions for the 1--component of the electric field in S, in terms
   of quantites defined in S':
   \begin{equation}  
 E^1_{\pm}  = \gamma^2(1\pm\beta^2) \partial'^1 A'^0+  \gamma^2(\beta^2 \pm 1) \partial'^0 A'^1
                  + \gamma^2 \beta(1\pm 1)( \partial'^0 A'^0+  \partial'^1 A'^1)
   \end{equation}
 Only the choice $E^1 \equiv E^1_-$ yields a covariant definition of the electric field.
  In this case, using Eqns(2.5) and (2.6), Eqn(3.2) simplifies to:
 \begin{equation}
   E^1 = \partial'^1 A'^0-\partial'^0 A'^1 =  E'^1
   \end{equation}
   Which expresses the well-known invariance of the longitudinal component of the electric 
   field under the LT.
   \par Thus, from rotational invariance, the general covariant definition of the electric field is:
  \begin{equation}
   E^i  = \partial^i A^0 - \partial^0 A^i
   \end{equation}
   This is the `additional assumption', mentioned by Jackson in the passage quoted above, that
  is necessary, in the present case, to derive the Lorentz force law. However, as written, it
  concerns only the physical properties of the electric field: the magnetic field concept has not
  yet been introduced. A further {\it a posteriori} justification of Eqn(3.4) will be given after
   derivation of the Lorentz force law. Here it is simply noted that, if the spatial part
  of the 4--vector potential is time-independent, Eqn(3.4) reduces to the usual electrostatic
  definition of the electric field.  
 \par The force $\vec{F}'$ on an electric charge $q$ at rest in the frame S' is given by the
  definition of the electric field, and Eqn(3.4) as:
    \begin{equation}
  F'^i = q(\partial'^i A'^0- \partial'^0 A'^i)
   \end{equation} 
   Equations analagous to (2.24) may be written relating $A'$ and $\partial'$ to the
   corresponding quantities in the frame S moving along the x' axis with velocity $-v$
   relative to S':
   \begin{eqnarray}
  \partial'^0 & = & U(\beta)\cdot  \partial \\
  A'^0 & = & U(\beta)\cdot A 
  \end{eqnarray} 
   Substituting (3.6) and (3.7) in (3.5) gives:
      \begin{equation}
   F'^i = q \left[ \partial'^i( U(\beta)\cdot A)-(U(\beta)\cdot \partial) A'^i \right]
   \end{equation}
   This equation expresses a linear relationship between $F'^i$, $\partial'^i$ and
  $A'^i$. Since the coefficients of the relation are Lorentz invariant,
   the same formula is valid in any inertial frame, in particular, in the frame S.
   Hence:
       \begin{equation}
   F^i = q \left[ \partial^i( U(\beta)\cdot A)-( U(\beta)\cdot  \partial) A^i \right]
   \end{equation}
   This equation gives, in 4--vector notation, a spatial component of the Lorentz force on the
   charge $q$ in the frame S, and so completes the derivation.
   \par To express the Lorentz force formula in the more familiar 3-vector notation,
   it is convenient to introduce the relativistic generalisation of Newton's Second Law
   ~\cite{Goldstein}:
   \begin{equation}
    \frac{dP}{d \tau} = F
   \end{equation}
    where $F$ is the 4-vector force and $\tau = t'$ is the proper time (in S') that is related
      to the time $t$ in S by the relativistic time dilatation formula:
     $d t = \gamma d \tau$.
    This gives, with Eqn(3.9) and (3.10):
   \begin{eqnarray}
     \frac{dP^i}{d \tau} & = &  \gamma \frac{dP^i}{d t} 
      \nonumber  \\
       & = & q (\partial^i A^{\alpha}-\partial^{\alpha} A^i) U(\beta)_{\alpha}
       \nonumber  \\
       & = &  \gamma q \left[\partial^i A^0-\partial^0 A^i-\beta_j
    (\partial^i A^j-\partial^j A^i)-\beta_k (\partial^i A^k-\partial^k A^i)   \right]
   \end{eqnarray}
    Introducing now the {\it magnetic field} according to the definition~\cite{Epsilon}:
   \begin{equation}
     B^k  \equiv  -\epsilon_{ijk}(\partial^i A^j-\partial^j A^i)= (\vec{\nabla}\times\vec{A})^k 
   \end{equation}
    enables Eqn(3.11) to be written in the compact form:
   \begin{equation}
    \frac{dP^i}{d t} =  q \left[E^i+ \beta_jB^k-\beta_k B^j \right] =
   q \left[E^i+ (\vec{\beta}\times \vec{B})^i \right] 
   \end{equation}
    so that, in 3--vector notation, the Lorentz force law is: 
  \begin{equation}
    \frac{d\vec{p}}{d t} =  mc \frac{d \gamma \vec{\beta}}{d t} = q [\vec{E}+\vec{\beta}\times \vec{B}]
   \end{equation}
  \par   Writing Eqn(3.4) in 3--vector notation and performing vector multiplication
   of both sides by the differential operator $\vec{\nabla}$ gives:
  \begin{equation}
 \vec{\nabla} \times \vec{E} = (\vec{\nabla} \times \vec{\nabla})A^0- \partial^0(\vec{\nabla}\times \vec{A})
  = -\frac{\partial \vec{B}}{\partial t}
    \end{equation}
  where Eqn(3.12) has been used. Eqn(3.15) is just the Faraday-Lenz induction law, i.e. the
   magnetodynamic Maxwell equation. This is only apparent, however, once the `magnetic field'
   concept of Eqn(3.12) has been introduced. Thus the initial hypothesis, Eqn(3.4), is actually a Maxwell
  equation. This is the {\it a posteriori} justification, mentioned above, for this covariant 
   definition of the electric field.
   \par It is common in discussions of electromagnetism to introduce the second rank
   electromagnetic field tensor, $F^{\mu \nu}$ according to the definition:
  \begin{equation}
 F^{\mu \nu} \equiv \partial^{\mu} A^{\nu}- \partial^{\nu} A^{\mu}
   \end{equation}
   in terms of which, the electric and magnetic fields are defined as:
  \begin{eqnarray}
   E^i &\equiv& F^{i0} \\
  B^k &\equiv&  -\epsilon_{ijk} F^{ij}
  \end{eqnarray}
  From the point of view adopted in the present paper both the electromagnetic field
   tensor and the electric and magnetic fields themselves are auxiliary quantities introduced
   only for mathematical convenience, in order to write the equations of electromagnetism in
    a compact way.
    Since all these quantities are completly defined by the 4--vector
   potential, it is the latter quantity that encodes all the relevant physical information
   on any electrodynamic problem~\cite{LCL2}. This position is contrary to that commonly taken in
  the literature and texbooks where it is often claimed that only the electric and magnetic fields
   have physical significance, while the 4--vector potential is only a convenient mathematical
  tool. For example R\"{o}hrlich~\cite{Rohrlich} makes the statement:
   \par{ \sl These functions ($\phi$ and $\vec{A}$) known as \underline{ potentials} have no physical
   meaning and are introduced solely for the purpose of mathematical simplification of the equations.}
   \par In fact, as shown above (compare Eqns(3.11) and (3.13)) it is the introduction of the electric
    and magnetic fields that enable the Lorentz force equation to be written in a simple manner!
    In other cases (e.g. Maxwell's equations) simpler expessions may be written in terms of the
    4--vector potential. The quantum theory, quantum electrodynamics, that underlies classical
   electromagnetism, requires the introduction the 4--vector photon field ${\cal A}^{\mu}$ in order\
   to specify the minimal interaction that provides the dynamical basis of the theory. Similarly,
     the introduction  of $A^{\mu}$ is necessary for the Lagrangian formulation of classical
     electromagnetism.
     It makes no sense, therefore, to argue that a physical concept of such fundamental
   importance has `no physical meaning'.    
   \par The initial postulate used here to derive the Lorentz force law is Eqn(3.4), which
     contains, explicitly, the electrostatic force law and, implicitly, the 
    Faraday-Lenz induction law. The actual form of the electrostatic force law (Coulomb's inverse
    square law) is not invoked, suggesting that the Lorentz force law may be of greater
    generality. On the assumption of Eqn(3.4) (which has been demonstrated to be the only
    possible covariant definition of the electric field), 
    the existence of the `magnetic field', the `electromagnetic field tensor', and finally
    the Lorentz force law itself have all been derived, without further assumptions, by use
    of the invariant formulation of the Lorentz transformation.
     \par It is instructive to compare the derivation of the Lorentz force law given in
     the present paper with that of Reference~\cite{LCL} based on the relativistic transformation
     properties of the Coulomb force 3--vector. Coulomb's law is not used in the present paper. On the
     other hand, Reference~\cite{LCL} makes no use of the 4--vector potential concept, which
     is essential for the derivation presented here. This demonstrates an interesting redundancy
     among the fundamental physical concepts of classical electromagnetism.
    \par In Reference~\cite{Schwartz}, Eqns(3.4), (3.12) and (3.16) were all introduced
     as {\it a priori} initial postulates without further justification.
     In fact, Schwartz gave the following
     explanation for his introduction of Eqn(3.16)~\cite{Shwartzgod}:
    \par  {\sl So far everything we have done has been entirely deductive, making use only of
        Coulomb's law, conservation of charge under Lorentz transformation and Lorentz
       invariance for our physical laws. We have now come to the end of this deductive 
      path. At this point when the laws were being written, God had to make a decision.
       In general there are 16 components of a second-rank tensor in four dimensions.
       However, in analogy to three dimensions we can make a major simplification 
     by choosing the completely antisymmetric tensor to represent our field quantities.
    Then we would have only 6 independent components instead of the possible 16. Under Lorentz
    transformation the tensor would remain antisymmetric and we would never have need for more
  than six independent components. Appreciating this, and having a deep aversion
   to useless complication, God naturally chose the antsymmetric tensor as His medium
    of expression.}
   \par Actually it is possible that God may have previously invented the 4--vector potential
   and special relativity, which lead, as shown above, to Eqn(3.4) as the only possible covariant
   definition of the
   electric field. As also shown in the present paper, the existence of the remaining elements
   of the antisymmetric field tensor, containing the magnetic field, then follow from
    special relativity alone.
    Schwartz derived the Lorentz 
     force law, as in Einstein's original Special Relativity paper~\cite{Einstein},
     by  Lorentz transformation of the electric field, from the rest frame of the test charge, to 
    one in which it is in motion. This requires that the magnetic field concept has previously
    been introduced as well as knowledge of the  Lorentz transformation laws of the electric
    and magnetic fields.
    \par In the chapter devoted to special relativity in Jackson's
    book~\cite{Jack2} the Lorentz force law is simply stated, without any derivation, as are also the
    defining equations of the electric and magnetic fields and the electromagnetic
   field tensor just mentioned. No emphasis is 
    therefore placed on the fundamental importance of the 4--vector potential in the relativistic
     description of electromagnetism.
   \par In order to 
    treat, in a similar manner, the electromagnetic and gravitational fields, the discussion
   in Misner Thorne and Wheeler~\cite{MTW} is largely centered on the properties of the tensor
    $F^{\mu \nu}$.
   Again the Lorentz force equation is introduced, in the spirit of the passage quoted above,
   without any derivation or discussion of its meaning. The defining equations of the electric
   and magnetic fields and $F^{\mu \nu}$, in terms of $A^{\mu}$, appear only in the eighteenth
    exercise of the relevant chapter. The main contents of the chapter on the electromagnetic
    field are an extended discussion of purely mathematical tensor manipulations
    that obscure the essential simplicity of electromagnetism
    when formulated in terms of the 4--vector potential.
     \par In contrast to References~\cite{Schwartz, Jack2, MTW}, in the derivation of the Lorentz 
     force law and the magnetic field presented here, the only initial assumption, apart from
    the validity of special relativity, is the chosen definition, Eqn(3.4), of the electric field
     in terms of the 4--vector potential $A^{\mu}$, which is the only  covariant one. Thus, a more
    fundamental description of electromagnetism than that provided by the electric and magnetic field
    concepts is indeed possible, contrary to the opinion expressed in the
    passage from Misner Thorne and Wheeler quoted above. 

 \SECTION{\bf{The time component of Newton's Second Law in Electrodynamics}} 
  
  Space-time exchange symmetry~\cite{JHF} states that physical laws in flat space are invariant 
  with respect to the exchange of the space and time components of 4-vectors.
   For example, the LT of time, Eqn(2.4), is obtained from that for space, Eqn(2.1), 
   by applying the space-time exchange (STE) operations: $x_0 \leftrightarrow x$,
   $x'_0 \leftrightarrow x'$. In
  the present case, application of the STE operation to the
  spatial component of the Lorentz force equation in the second line of Eqn(3.11)
  leads to the relation:
   \begin{eqnarray}
     \frac{dP^0}{d \tau} & = & \frac{\gamma}{c} \frac{dP^0}{d t} = 
     q (\partial^0 A^{\alpha}-\partial^{\alpha} A^0) U(\beta)_{\alpha}
      \nonumber  \\
       & = & -q E^i U(\beta)_i = \gamma q \frac{\vec{E}\cdot \vec{v}}{c}
   \end{eqnarray}    
 where Eqns(2.5) and (3.4) and the following
 properties of the STE operation~\cite{JHF} have been used:
  \begin{eqnarray}
  \partial^0 & \leftrightarrow & - \partial^i
     \\
  A^0 &\leftrightarrow& - A^i
     \\
   C \cdot D &\leftrightarrow& - C \cdot D
     \end{eqnarray} 
  Eqn(4.1) yields an expression for the time derivative of the relativistic
  energy, ${\cal E}= P^0$ :
 \begin{equation}  
     \frac{d{\cal E}}{ d t} = q \vec{E}\cdot \vec{v} = q  \vec{E}\cdot \frac{d \vec{x}}{dt}
 \end{equation}
 Integration of Eqn(4.5) gives the equation of energy conservation for a particle
  moving from an initial position, $\vec{x}_I$, to a final
   position, $\vec{x}_F$,  under the influence of electromagnetic forces:
   \begin{equation} 
    \int_{{\cal E}_I}^{{\cal E}_F} d {\cal E} = q  \int_{\vec{x_I}}^{\vec{x_F}} \vec{E}\cdot d \vec{x}
    \end{equation} 
  Thus work is done on the moving charge only by the electric field. This is also evident
  from the Lorentz force equation, (3.14), since the magnetic force
   $\simeq \vec{\beta}\times \vec{B}$ is perpendicular to the velocity vector, so that
    no work is performed by the magnetic field. A corollary is that the relativistic
   energy (and hence the magnitude of the velocity) of a charged particle moving
   in a constant magnetic field is a constant of the motion. Of course, Eqn(4.5)
   may also be derived directly from the Lorentz force law, so that the time
   component of the relativistic generalisation of Newton's Second Law, Eqn(4.1),
   contains no physical information not already contained in the spatial components.
   This is related to the fact that, as demonstrated in Eqns(2.25) and (2.26),
   the spatial and temporal components of the energy-momentum 4--vector
   are not independent physical quantities.
 
    \par{\bf Acknowledgements}
     \par I should like to thank O.L. de Lange for asking the question
     whose answer, presented in Section 4, was the  original motivation for
     the writing of this paper, and an anonymous referee of an earlier version
     of this paper for informing me of related material, in the books of
     Jackson and Misner, Thorne and Wheeler, which is discussed in some detail
     in this version. 
\pagebreak
 

\begin{thebibliography}{99}
\bibitem{JHF1}
 Actually, a careful examination of the derivation of Amp\`{e}re's from the Gauss
 law of electrostatics in Reference[3] shows that, although Eqn(3.4) of the
  present paper is a necessary initial assumption, the definition of the magnetic
  field in terms of the spatial derivatives of the 4--vector potential occurs naturally
 in the course of the derivation (see Eqns(5.16) and (5.17) of Reference[3]) so it 
  is not necessary to assume, at the outset, the expression for the spatial components of
  the electromagnetic field tensor as given by Eqn(5.1) of Reference[3].
\bibitem{Fahnline}
 This should not be confused with a manifestly {\it covariant} 
  expression for the LT, where it is written as a linear
  4-vector relation with Lorentz-invariant coefficients, as in:
  D.E.Fahnline,
 Am. J. Phys. {\bf 50} 818 (1982).
\bibitem{Defns}
  A time-like metric is used for 4-vector products with the components
   of a 4--vector, $W$, defined as:  \[ W_t = W^0 =W_0,~~
  W_{x,y,z} = W^{1,2,3} = -W_{1,2,3}\]  and an implied summation over repeated
  contravariant (upper)  and covariant (lower) indices. Repeated Greek
 indices are summed from 0 to 3, repeated Roman ones from 1 to 3. Also
 \[ \partial^{\mu} \equiv (\frac{\partial}{\partial x^0}; -\frac{\partial}{\partial x^1},
 -\frac{\partial}{\partial x^2}, -\frac{\partial}{\partial x^3})
   = (\partial^0; -\vec{\nabla}) \]
\bibitem{LCL2}
 The explicit form of $A^{\mu}$, as derived from Coulomb's law, is given in standard textbooks 
   on classical electrodynamics. For example, in Reference[13], it is to be found in
   Eqns(17-51) and (17-52). $A^{\mu}$ is actually proportional to the 4-vector velocity, $V$, of
    the charged particle that is the source of the electromagnetic field.
\end{thebibliography}
\end{document}